\documentclass[aps,twocolumn,showpacs]{revtex4-1}
\usepackage{float}
\usepackage{makeidx}
\usepackage{color}
\usepackage{amsmath}
\usepackage{braket}
\usepackage{amsthm}
\usepackage{amssymb}
\usepackage{amsfonts}
\usepackage{graphicx,subfigure}
\usepackage{txfonts}
\usepackage{ulem}
\usepackage{mathtools}
\usepackage{bm}
\usepackage{hyperref}
\usepackage{lineno}
\usepackage{natbib}
\usepackage{comment}
\usepackage{amsmath,amssymb,graphics,epsfig,color,times,bbm}

\newcommand{\vect}[1]{\bm{#1}}

\newcommand{\Eins}{\ensuremath{\mathbbm 1}}
\newcommand{\mean}[1]{\ensuremath{\big\langle #1 \big\rangle}}

\newcommand{\be}{\begin{equation}}
\newcommand{\ee}{\end{equation}}
\newcommand{\beq}{\begin{eqnarray}}
\newcommand{\eeq}{\end{eqnarray}}

\begin{document}

%\title{Scalable entangled network of Mach-Zehnder interferometers for distributed quantum sensing} 
\title{Scalable Network of Mach-Zehnder Interferometers with a Single Entangled Resource}
\author{Zhihui Yan$^{1,2}$}
\thanks{These authors contributed equally to this work.}
\author{Yanni Feng$^{1}$}
\thanks{These authors contributed equally to this work.}
\author{Luca Pezze$^{3}$}
\thanks{These authors contributed equally to this work.}
\author{Zhaoqing Zeng$^{1}$}
\author{Jingxu Ma$^{1}$}
\author{Xiaoyu Zhou$^{1}$}
\author{Augusto Smerzi$^{3}$}
\email{augusto.smerzi@ino.it}
\author{Xiaojun Jia$^{1,2}$}
\email{jiaxj@sxu.edu.cn}
\author{Kunchi Peng$^{1,2}$}
\affiliation{$^{1}$State Key Laboratory of Quantum Optics Technologies and Devices, Institute of Opto-Electronics, Shanxi University, Taiyuan, 030006, People's Republic of China \\
$^{2}$Collaborative Innovation Center of Extreme Optics, Shanxi University, Taiyuan 030006, People's Republic of China\\
$^{3}$ INO-CNR and LENS, Largo Enrico Fermi 6, Firenze, 50125, Italy}

\begin{abstract}

Distributed quantum sensing exploits entanglement to enhance the estimation of multiple parameters across a network of spatially-separated sensors, achieving sensitivities beyond the classical limit. 
Potential applications cover a plethora of technologies, from precision navigation to biomedical imaging and environmental monitoring.
However, practical implementations are challenged by the complex optimal distribution of entanglement throughout the sensing nodes, which affects scalability and robustness.
Here we demonstrate a reconfigurable network of Mach–Zehnder interferometers entangled via a single shared squeezed-vacuum resource. 
We achieve joint noise suppression of $4.36 \pm 0.35$ dB below the standard quantum limit at the phase-uncertainty level of $10^{-9}$ .
Furthermore, after full optimization in the low-intensity regime, we demonstrate a crossover from the standard quantum limit to the Heisenberg limit. 
The network estimates arbitrary linear combinations of phases, saturates the quantum Cramér-Rao bound in the ideal case, remains robust under realistic photon losses, and scales favorably with the number of sensors. 

\end{abstract}

\maketitle

Quantum interferometers are the most sensitive measurement devices in modern science and technology, with applications ranging from gravitational-wave detection~\cite{AbbottPRL2016} to nanoscale imaging~\cite{CasacioNATURE2021}.
A foundational example is the Mach-Zehnder interferometer (MZI), which serves as a key platform in both fundamental studies and practical implementations~\cite{PirandolaNATPHOT2018}.
The MZI offers enhanced phase estimation by exploiting high-intensity optical fields and common-mode noise rejection, with a precision fundamentally limited by quantum fluctuations~\cite{SchnabelNATCOMM2010, LawrieACSP2019, PolinoAVSQS2020}.
While conventional MZIs operate at the standard quantum limit (SQL)~\cite{PezzePRL2007}, sub-SQL sensitivities have been demonstrated by injecting nonclassical states -- such as squeezed or entangled light~\cite{CavesPRD1981, XiaoPRL1987, HollandPRL1993, PezzePRL2008, JooPRL2011, LangPRL2013} -- into the interferometer.
These advancements have been realized in proof-of-principle experiments~\cite{NagataSCIENCE2007, AfekSCIENCE2010, SlussarenkoNATPHOT2017, DaryanooshNATCOMM2018, ZuoPRL2020, NielsenPRL2023} as well as in large-scale observatories such as Advanced LIGO and Virgo~\cite{GodaNATPHYS2008, AbadieNATPHYS2011, AasiNATPHOT2013, TsePRL2019, AcernesePRL2019, JiaSCIENCE2024}.

Most quantum-enhanced interferometric implementations remain focused on estimating a single phase shift. 
However, a wide range of emerging applications -- including biological imaging~\cite{Juan, Hoffmann, DefienneNATPHOT2024}, inertial navigation~\cite{GravePRAPP2020} and environmental monitoring~\cite{G2022} -- requires the simultaneous estimation of multiple phases across a network of distributed sensing nodes.
To address this need, entanglement-enhanced distributed quantum sensing (DQS) is rapidly gaining prominence as a key quantum technology~\cite{PezzeARXIV2025}. 

%%%%%%%%%%%%%%%%%%%%%%%%%%%%
%% Figure 1
%%%%%%%%%%%%%%%%%%%%%%%%%%%%
\begin{figure}[bh!]
\centering
\includegraphics[width=\columnwidth]{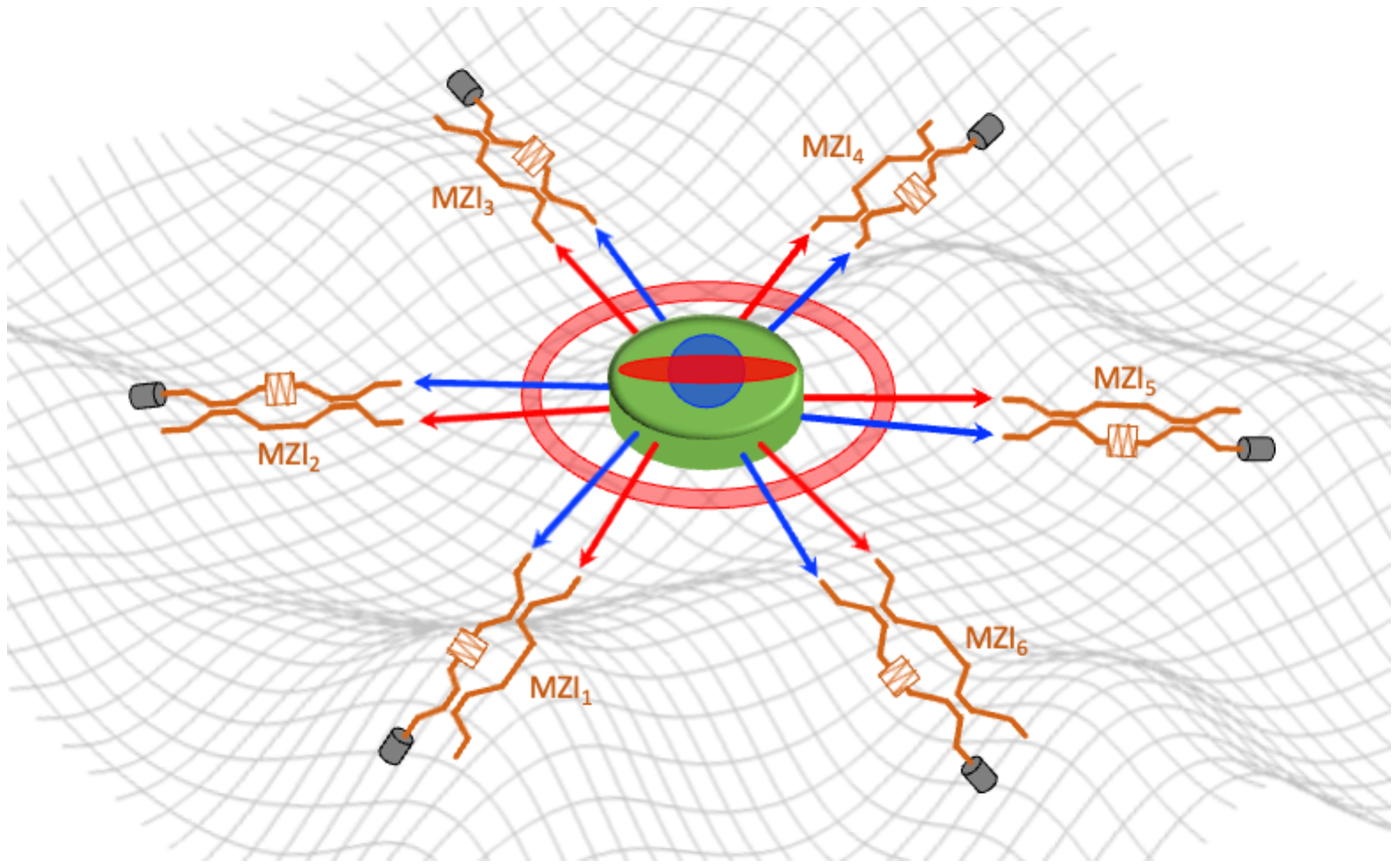}
\caption{\textbf{Distributed quantum sensing with a scalable entangled network of Mach-Zehnder interferometers.} 
In this schematic architecture, the central node (green cylinder) is responsible for generating and distributing mode entanglement across spatially separated MZIs (orange).
The node contains a squeezer (red ellipse), a displacer (blue disc), and a quantum circuit.
Each local MZI receives two input modes: a coherent mode (blue arrow) and an entangled mode (red arrow): mode entanglement generated by the multimode splitting of a single squeezed-vacuum state is represented by the red halo.
Within each MZI, the two inputs are combined at a beam splitter, acquire a relative phase shift (the zig-zag lines indicating multipass phase amplification), and are then recombined at a second beam splitter before detection.
Measurement outcomes are collected from a single output port of each MZI.
The wavy background schematically represents an inhomogeneous field.
This field couples to the interferometers by inducing local relative phase shifts between the two sensing arms.
}
\label{Figure1}
\end{figure}
%%%%%%%%%%%%%%%%%%%%%%%%%%%%
%%%%%%%%%%%%%%%%%%%%%%%%%%%%
%%%%%%%%%%%%%%%%%%%%%%%%%%%%

%%%%%%%%%%%%%%%%%%%%%%%%%%%%
%% Figure 2
%%%%%%%%%%%%%%%%%%%%%%%%%%%%
\begin{figure*}[th!]
\centering
\includegraphics[width=\textwidth]{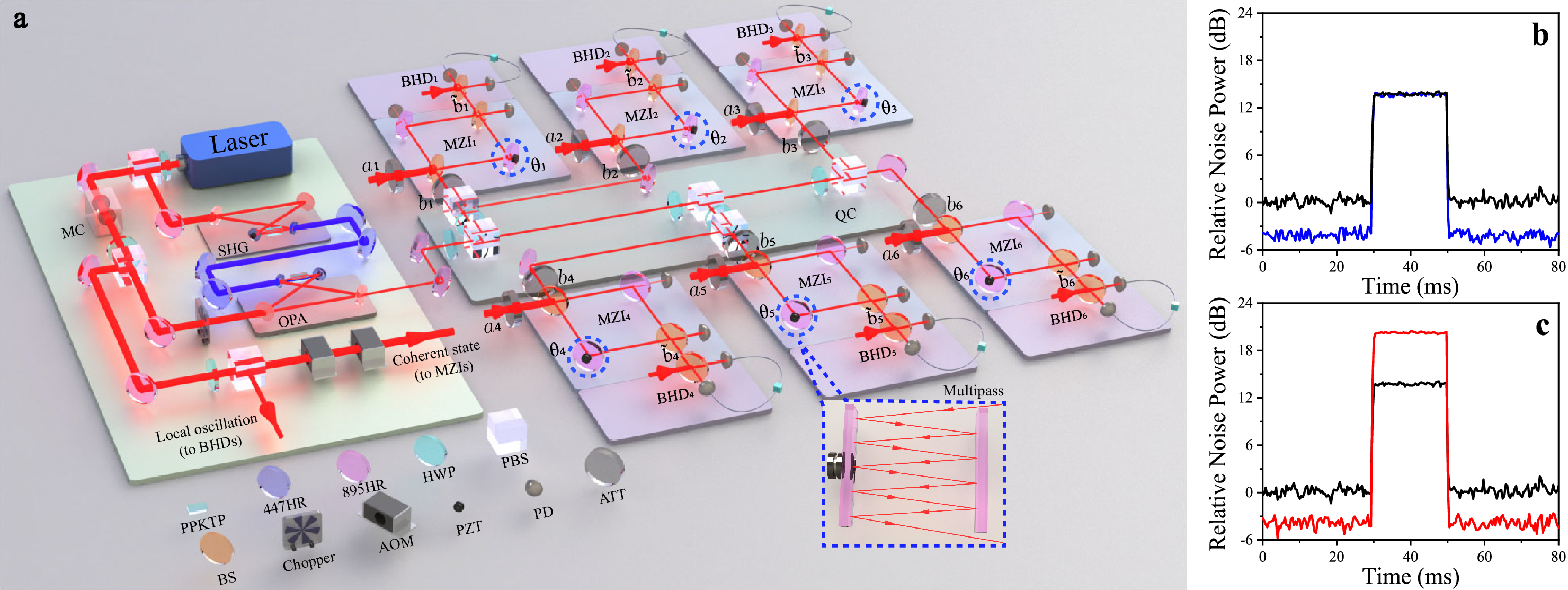}
\caption{\textbf{Experimental realization of an entangled network of quantum interferometers.}
\textbf{a} Schematic of the experimental apparatus. 
A single squeezed-vacuum state is generated by an optical parametric amplifier (OPA) and evenly distributed via a quantum circuit (QC) composed of polarization beam splitters. The remaining $d - 1$ inputs to the QC are vacuum states (with $d = 6$ in this implementation). Linear splitting of the squeezed input generates multipartite entanglement across the $d$ output modes, labeled $b_1, ..., b_d$.
Each output mode $b_j$ is injected into the $j$th MZI (thin arrow), while the second input port $a_j$ receives a coherent state (thick arrow). Each MZI comprises a balanced beam splitter, a phase interrogation region where the beam undergoes $K$ passes through the sample, and a final balanced beam splitter. The phase shift accrued in a single-pass through the $j$th interferometer is denoted by $\theta_j$.
The output mode $\tilde{b}_j$ (thin line) is analyzed by a local balanced homodyne detection (BHD), using an external local oscillator (thick arrow). Multiphase estimation is achieved by jointly processing all BHD measurement outcomes.
\textbf{(Inset)} Detail of the multipass interaction. PPKTP: periodically poled KTiOPO$_4$; HR: high-reflectivity mirror; HWP: half-wave plate; PBS: polarization beam splitter; BS: beam splitter; AOM: acousto-optic modulator; PZT: piezoelectric transducer; PD: photodetector; ATT: optical attenuator; MC: mode cleaner; SHG: second harmonic generator.
Panels \textbf{b}
and \textbf{c} show the joint relative noise powers of the phase average $(%
\protect\theta_1+...+\protect\theta_d)/d$. The black solid line corresponds
to the conventional network of MZI (namely, with no squeezing and single-pass interaction, $K=1$). It is compared with the noise powers obtained with
the entangled network of MZIs (colored line) for single-pass (panel \textbf{b%
}) and five-pass (panel \textbf{c}) interactions. 
In spectrum analyzer, analyzing frequency is 4 MHz; resolution bandwidth is 100 kHz; video bandwidth is 1 kHz.
}
\label{fig2}
\end{figure*}
%%%%%%%%%%%%%%%%%%%%%%%%%%%%
%%%%%%%%%%%%%%%%%%%%%%%%%%%%
%%%%%%%%%%%%%%%%%%%%%%%%%%%%

In this manuscript, we realize experimentally a novel DQS optical architecture consisting of a scalable and versatile network of MZIs. 
The multiparameter sensor is schematically shown in Fig.~\ref{Figure1} and builds on recent theoretical proposals~\cite{11, PezzeARXIV}.
The scheme estimates arbitrary linear combinations of the relative phase shift in each MZI through optimal local measurements.
The splitting of a single Gaussian squeezed-vacuum state generates entanglement among multiple modes. 
Each entangled mode enters one input port of a MZI, while a coherent state enters its second port.
The hybrid quantum-classical nature of the probe state is a distinguishing trait of our scheme.
Specifically, we exploit intense coherent states to reach phase uncertainties on the order of $10^{-9}$ with the total input power of 9.6 $mW$, achieving more than 4 dB below the SQL thanks to the distributed squeezed vacuum.
We report absolute quantum-enhanced phase sensitivities that are orders of magnitude beyond those achieved in the current literature on multiparameter phase sensing.
We further extend the approach to the opposite regime of low photon-intensity and optimize the distribution of photons between the split squeezed vacuum and the coherent states.
In particular, we address a crossover regime between the SQL -- which is proved optimal in the limit of vanishing intensities -- and the Heisenberg limit, and demonstrate a sub-SQL sensitivity with respect to the total average photon number in the sensor.
Unlike prior multiphase estimation schemes~\cite{LiuNATPHOT2021, CiminiAP2023, HongNATCOMM2021, KimNATCOMM2024, ZhaoPRX2021}, our approach makes use of Gaussian states which prove inherently robust. 
Furthermore, our sensor network exploits the notable practical advantages of MZIs -- offering noise-resilient differential measurements, lack of external phase reference and reduced sensitivity to local oscillator and alignment noise -- for multi-phase estimation beyond homodyne~\cite{GuoNATPHYS2020} and quadrature~\cite{XiaPRL2020} sensing schemes.
We directly demonstrate the protocol’s resilience to photon losses and show that a single squeezed-vacuum source is sufficient to scale the sub-SQL sensing protocol to interferometric arrays of varying sizes -- up to six MZIs in our experiment.
Our results offer a practical path toward quantum-enhanced DQS, with potential applications across a broad range of scientific and technological platforms.\\

\textbf{The entangled sensor network.} \\

Figure~\ref{fig2}(a) presents the experimental apparatus.
A squeezed-vacuum state is generated by an optical parametric amplifier (OPA) and split through a quantum circuit (QC) composed of a cascade of beam splitters.
The remaining inputs of the QC are vacuum states.
The configurable linear optical splitting \cite{11,PezzeARXIV} generates
multipartite entanglement among the output modes $b_1, ..., b_d$ (with $d = 6$ in our implementation).
The sensor network is composed of $d$ MZIs.
The entangled mode $b_j$, output of the QC, is injected into one input port of the $j$th MZI. 
This establishes nonclassical correlations across the network that suppress joint quantum noise in the multiparameter estimation process.
The remaining input ports $a_1, \dots, a_d$ of the MZIs are fed with coherent states to enhance the phase sensitivity via increased photon number. 
To amplify the phase signal, we incorporate a multipass configuration using two high-reflectivity mirrors within each interferometer arm. 
One mirror interfaces with the sample under interrogation, while the second mirror facilitates $K$ back-and-forth reflections, resulting in a phase shift that scales linearly with $K$.
Phase information is extracted via balanced homodyne detection (BHD) on the output modes, specifically targeting the phase quadrature at the destructive interference ports. 
The results from each local measurement are jointly processed to estimate the relevant phase parameters.
Our goal is to estimate linear combinations 
\begin{equation} \label{combination}
\bm{\nu}^\top \bm{\theta} =\sum\limits_{j=1}^{d} \nu_{j}\theta _{j},
\end{equation}
where $\bm{\theta}= \{\theta_1,
..., \theta_d\}^\top$, $\theta_j$ is the relative phase shifts between the sensing modes in the $j$th MZI, and $\bm{\nu}=\{\nu_1, ..., \nu_d\}^\top$ is a vector of real weights.
By jointly processing the BHD signals, we obtain
\begin{equation}  \label{sensitivity}
\Delta^2 (\bm{\nu}^\top \bm{\theta}) = \bm{\nu}^{\top} \bm{C}^{-1} 
\bm{\Gamma} (\bm{C}^{\top})^{-1} \bm{\nu},
\end{equation}
as provided by error propagation~\cite{1}, where $\bm{C}_{ij} = \frac{\partial \langle \hat{Q}_i \rangle}{\partial \theta_j}$
and $\bm{\Gamma}_{ij} = \langle \hat{Q}_i \hat{Q}_j \rangle - \langle \hat{Q}_i \rangle
\langle \hat{Q}_j \rangle$ are $d\times d$ matrices, $\hat{Q}_j$ are the quadrature
operators measured by BHD, and the expectation values are computed on the
output state of the network. 

We evaluate Eq.~(\ref{sensitivity}) analytically and optimize it with 
%respect to the photon number of the $d$ coherent states as well as with 
respect to the splitting of the squeezed-vacuum state~\cite{supp}. 
The optimization gives 
\begin{equation}  \label{sensEnt}
\Delta^2 (\bm{\nu}^\top \bm{\theta})_{\mathrm{opt}} = \frac{e^{-2r}+ \Lambda%
}{K \bar{n}_T} \qquad {\rm for} \,\, \bar{n}_c \gg \bar{n}_s,
\end{equation}
where $r$ is the squeezing strength of the squeezed vacuum state, $\Lambda$ quantifies the photon losses ($\Lambda=0$ in the ideal case), $\bar{n}_c$ the total mean photon number of the $d$ coherent states, $\bar{n}_s$ the intensity of the squeezed vacuum, and $\bar{n}_T = \bar{n}_c + \bar{n}_s$. 
The multipass enhancement factor is $k = \mu K^2$, where $K$ is the number of multipass interaction and $\mu$ is multipass coefficient arising from different response of multipass interaction in a piezoelectric transducer (PZT).
In our case $\mu \approx 1/K$ within the experimental accessible range of six passes, giving an effective sensitivity enhancement $K$ as in Eq.~(\ref{sensEnt}).  
Photon losses are captured by the factor $\Lambda= 1/\eta-1$ where the total transmission efficiency is $\eta = \eta_{dis}\eta_{MZI} \eta_{m}^{2K-1}$.
Here, $\eta_{dis}$ is the distribution efficiency between the QC and MZI, $\eta_{MZI}$ is the internal efficiency of each interferometer and $\eta_m$ quantifies the losses in the multipass system.
The lossless case is recovered for unit transmission, namely $\eta=1$ (corresponding to $\Lambda=0$).
Overall, Eq.~(\ref{sensEnt}) shows that the SQL sensitivity for an ideal network of MZIs is recovered for $\bar{n}_s=0$ and $\Lambda=0$. 
The SQL is overcome when $\bar{n}_s\neq 0$, as long as losses are sufficiently small -- specifically, when $\Lambda < 1-e^{-2r}$. 

To further optimize the sensitivity for a fixed total photon number $\bar{n}_T$, we minimize Eq.~(\ref{sensEnt}) with respect to the squeezing parameter $r$~\cite{supp}.
We can distinguish three regimes~: 
\begin{equation} \label{optsqu}
\min_r~\Delta^2 (\bm{\nu}^\top \bm{\theta})_{\mathrm{opt}} =
\begin{cases}
\frac{1+\Lambda}{K\bar{n}_T} & {\rm for}\quad \bar{n}_T \ll  1 \\
\frac{1}{K\bar{n}_T^2} & {\rm for}\quad 1 \ll \bar{n}_T \lesssim 1/\Lambda \\
\frac{\Lambda}{K \bar{n}_T} & {\rm for}\quad  \bar{n}_T \gtrsim 1/\Lambda.
\end{cases}
\end{equation}
For sufficiently low losses, Eq.~(\ref{optsqu}) is characterized by the crossover from a SQL scaling with prefactor $1+\Lambda$, for low photon number, to the Heisenberg limit (HL), $1/(K \bar{n}_T^2)$.
The latter is achieved when half of the $\bar{n}_T$ particles are in the squeezed-vacuum state and the other half are in the $d$ coherent states, i.e. $\bar{n}_s = \bar{n}_c=\bar{n}_T/2$.
On the opposite, if $\bar{n}_T \ll 1$, it becomes convenient to allocate more particles in the coherent state as it has a narrower photon distribution than the squeezed-vacuum.
Finally, as generally expected for imperfect interferometers~\cite{EscherNATPHYS2011, RafalNATCOMM2012} ($\Lambda \neq 0$) the Heisenberg limit breaks down asymptotically in $\bar{n}_T$, and we find a SQL scaling with prefactor $\Lambda$. 

We can compare the performance of our entangled network with that of a separable strategy, in which $d$ independent squeezed vacuum states and $d$ coherent states are injected into the input ports, respectively, of independent MZIs~\cite{11,PezzeARXIV}.
When the squeezing strength $r$ is fixed, both strategies yield the same sensitivity for estimating a linear combination of the
parameters, $\bm{\nu}^\top \bm{\theta}$, as given by Eq.~(\ref{sensEnt}), see \cite{supp} for a detailed derivation.
However, the entangled strategy achieves this performance by using a single squeezed-vacuum resource, whereas the separable approach requires $d$ independent non-classical resources.
This represents a substantial reduction of experimental overhead and resource complexity.
Furthermore, when each squeezed-vacuum state in the separable scheme is independently optimized (even under lossy conditions) the sensitivity gain of the entangled strategy over the separable one is
\begin{equation} \label{gain}
G = 
\begin{cases}
\lVert \bm{\nu} \rVert_{2/3}^2 & \mathrm{for}~\bar{n}_T \lesssim \Lambda, \\ 
1 & \mathrm{for}~\bar{n}_T \gtrsim \Lambda,%
\end{cases}%
\end{equation}
where $\lVert \bm{\nu} \rVert_\kappa = (\sum_{j=1}^d \vert \nu_j \vert^\kappa)^{1/\kappa}$.
We have that $\lVert \bm{\nu} \rVert_{2/3}^2 \in [1, d]$. 
It is $\lVert \bm{\nu} \rVert_{2/3}^2 = 1$ when considering the estimation of a single phase shift, e.g., $\vect{\nu}_{\rm single} = \{1,0,...,0\}$.
Instead, $\lVert \bm{\nu} \rVert_{2/3}^2 = d$ for $\vect{\nu}_{\rm ave}
= \{1, 1,..., 1\}/d$, corresponding to the estimation of the average phase shift $\vect{\nu}^\top_{\rm ave} %
\bm{\theta} = \sum_{j=1}^d \theta_j/d$. 
In our experiments, we focus primarily on this latter scenario, as it provides the most stringent benchmark for DQS.

Finally, we emphasize that, at least in the noiseless case and for sufficiently large $\bar{n}_c$, the sensitivity Eq.(\ref{sensEnt}) achieved with the BHD measurements implemented in our setup saturates the quantum Cramér-Rao bound computed in Ref.~\cite{PezzeARXIV} (see Methods for a detailed discussion).
This validates the performance of our BHD.\\

%%%%%%%%%%%%%%%%%%%%%%%%%%%%
%% Figure 3
%%%%%%%%%%%%%%%%%%%%%%%%%%%%
\begin{figure*}[th!]
\centering
\includegraphics[width=\textwidth]{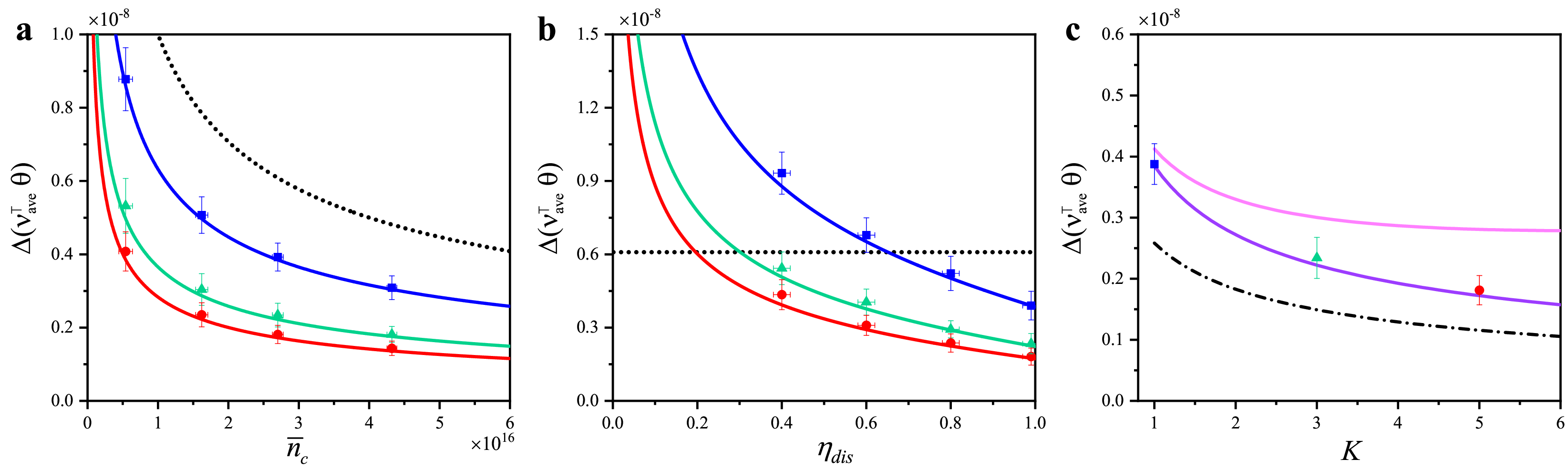}
\caption{\textbf{High-sensitivity regime of DQS.}
\textbf{a.} Phase uncertainty $\Delta (\vect{\nu}^\top_{\rm ave} \vect{\theta})$ as a function of the average photon number $\bar{n}_c$ in the coherent input states.
Blue squares, green triangles, and red circles denote experimental data for different values of the multipass number $K = 1$, 3, and 5, respectively.
The black dotted line shows the SQL, $1/\sqrt{\bar{n}_T}$, while the solid curves is the theoretical model from Eq.~(\ref{sensEnt}).
Here, the average photon number in the squeezed-vacuum state is $\bar{n}_s = 0.68 \pm 0.01$, and the system efficiencies are: $\eta_{\rm dis} = 99\%$, $\eta_{\rm MZI} = 89\%$, and $\eta_{m} = 99.99\%$.
\textbf{b.} Phase uncertainty $\Delta (\vect{\nu}^\top_{\rm ave} \vect{\theta})$ as a function of distribution efficiency $\eta_{\rm dis}$, for $\bar{n}_s = 0.68 \pm 0.01$ and $\bar{n}_c = (2.7 \pm 0.077) \times 10^{16}$. Symbols and lines are defined as in panel (a).
\textbf{c.} Phase uncertainty as a function of the multipass number $K$. The pink and purple curves represent sensitivities for $\eta_m = 95\%$ and $\eta_m = 99.99\%$, respectively. The black dashed-dotted line indicates the ideal sensitivity limit ($\eta = 1$). 
Notably, the improvement in sensitivity with increasing $K$ becomes evident for high detection efficiency, highlighting the practical potential of multipass schemes under realistic conditions.
}
\label{fig3}
\end{figure*}
%%%%%%%%%%%%%%%%%%%%%%%%%%%%
%%%%%%%%%%%%%%%%%%%%%%%%%%%%
%%%%%%%%%%%%%%%%%%%%%%%%%%%%

\textbf{Results.} \\

We experimentally explore the two different quantum sensing regimes discussed above, namely large and small total average photon numbers.
First, we consider the case $\bar{n}_c \gg \bar{n}_s$, where high-sensitivity estimation is achieved by using input coherent states of large photon number. 
The present regime mimics, for instance, that of a possible network of gravitational-wave detectors using large intensity laser fields at high frequency, when the radiation pressure noise at mirrors can be neglected. 
In this limit, Eq.~(\ref{sensEnt}) predicts a $\sim 1/(K\bar{n}_T)$ scaling of multiphase uncertainty with a prefactor that can be reduced below 1 (thus overcoming the SQL) thanks to a single and optimally-split squeezed-vacuum state, provided that losses are sufficiently low. 
After joint processing of the BHD signals, the measured noise powers of the entangled MZI network are shown in Fig.~\ref{fig2}(b) and (c) for single-pass and five-pass configurations, respectively.
The relative noise power reported in Fig.~\ref{fig2}(b,c) is normalized to the SQL, accounting for both signal and noise quadrature components to obtain the signal-to-noise ratio.
When the driven signals for PZTs are turned on between 30 and 50 ms, the phase related signal powers are measured during this interval. 
The noise power is measured during the remaining part of each 80 ms cycle when the PZTs are idle.
The mode-entangled resources reduce the joint noise by $4.36 \pm 0.35$ dB below the SQL for estimation of $\bm{\nu}_{\rm ave}^\top \bm{\theta}$. 
Furthermore, the $K=5$ pass interaction enhances the signal-to-noise ratio by $11.09 \pm 0.38$ dB compared to the conventional interferometric baseline (namely $K=1$).

A more systematic study is shown in Fig.~\ref{fig3}.
In Fig.~\ref{fig3}(a), we plot $\Delta (\bm{\nu}_{\rm ave}^\top \bm{\theta})$ as a function of $\bar{n}_c$ in the high photon number range $\bar{n}_T \sim \bar{n}_c \sim 10^{16}$.
The SQL performance of the sensor network (black dotted line for $K = 1$) improves with increasing photon number in the coherent input states and is further enhanced by incorporating a squeezed-vacuum state with $\bar{n}_s = 0.68$ (squeezing parameter $r = 0.75$), as shown by the blue squares.
The combined effect of squeezing and multipass enhancement is illustrated by the green triangles ($K = 3$) and red circles ($K = 5$). 
The experimental data are well reproduced by the theoretical line, Eq.~(\ref{sensEnt}), including a total transmission efficiency of $\eta = 88\%$. 
The phase change is caused by PZT vibration, and thus the standard deviation in the phase can reach 1.4$\times $10$^{-9}$ with the total input coherent power of 9.6 $mW$.

The robustness of the entangled MZI network to distribution losses is investigated in Fig.~\ref{fig3}(b). 
Experimental data were collected by tuning the distribution efficiency $\eta_{\mathrm{dis}}$, which was adjusted via the angle of half-wave plates in the polarization beam splitters (PBSs).
For single-pass operation ($K=1$), the sensitivity of the entangled network (blue squares and line) beats the SQL (horizontal dotted black line) for $\eta_{\mathrm{dis}} \gtrsim 65\%$. 
For the multipass scheme, (green triangles and line for $K=3$ and red circles and line for $K=5$), the robustness to loss can be enhanced with respect to the SQL. In particular, for five-pass interaction $K=5$, the entangled network outperforms the conventional sensing configuration for $\eta_{\mathrm{dis}} \gtrsim 20$$\%$.

Figure~\ref{fig3}(c) presents $\Delta (\bm{\nu}_{\rm ave}^\top \bm{\theta})$ as a function of the number of multipass interactions $K$, for fixed values of $\bar{n}_s$ and $\bar{n}_c$.
The performance of the sensor is characterized by two competing effects.
On the one hand multipass interactions between high-reflectivity mirrors amplify the phase signal, thereby improving overall sensitivity. 
On the other hand, the increased optical path length introduces vacuum noise via mirror losses, presenting a trade-off between signal enhancement and noise degradation. 
Thus, the benefits of increased $K$ depend on maintaining high mirror efficiency.
In our setup, the multipass interaction efficiency is extremely high ($\eta_m = 99.99\%$), while the combined distribution and interferometer efficiencies are $\eta_{\rm dis} \eta_{\rm MZI} = 88\%$. 
As a result, the sensitivity improvement is clearly observed even for a relatively moderate number of passes.

\begin{figure}[t!]
\centering
\includegraphics[width=0.85\columnwidth]{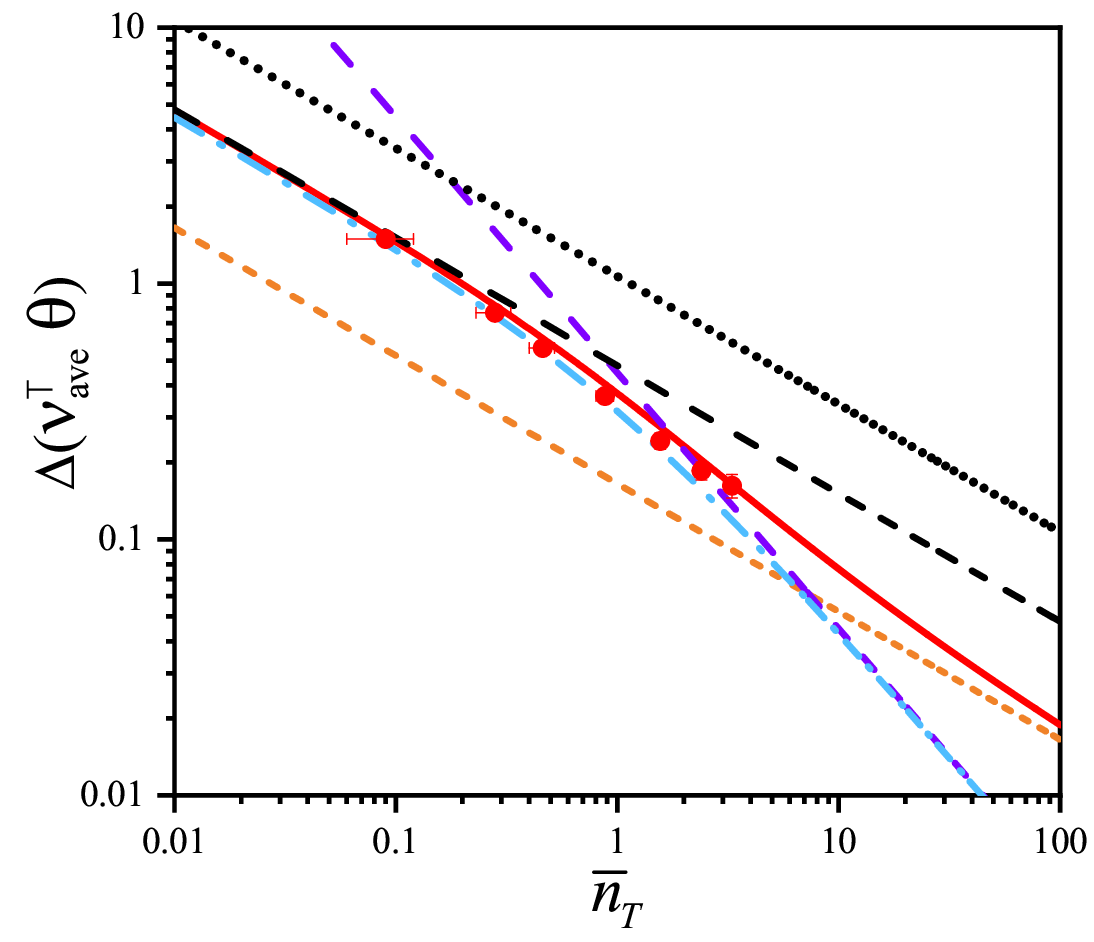}
\caption{\textbf{Optimized sensitivity of the entangled sensor network.} Multiphase sensitivity $\Delta(\bm{\nu}_{\rm ave}^\top \bm{\theta})$ as a function of total average photon number $\bar{n}_T$, with photon resources optimally allocated between squeezed and coherent components.
Dots represent experimental results obtained for $K = 5$.
The corresponding total photon numbers are $\bar{n}_T = 0.09$, $0.28$, $0.46$, $0.88$, $1.56$, $2.4$, and $3.29$, with associated squeezed-vacuum photon numbers are $\bar{n}_s = 0.006$, $0.04$, $0.09$, $0.21$, $0.42$, $0.68$, and $0.93$. 
The experimental data closely follow the theoretical prediction (red solid line) corresponding to a numerical optimization of Eq.(\ref{sensEnt}) with efficiencies of $\eta_{\rm dis} = 99\%$, $\eta_{\rm MZI} = 89\%$, and $\eta_m = 99.99\%$, each independently calibrated.
For $\bar{n}_T \gg 1$ the red line converges to $\sqrt{\Lambda /(K \bar{n}_T)}$ (orange dashed line), while for $\bar{n}_T \ll 1$ it saturates $\sqrt{(1+\Lambda)/( K\bar{n}_T)}$ (black dashed line).
The light blue dashed line is the optimized Eq.(\ref{sensEnt}) in the lossless case ($\Lambda=0$): it saturates the HL $1/(\sqrt{K}\bar{n}_T)$ for $\bar{n}_T \gg 1$ (violet dashed line).
Finally, the dotted line is the SQL for $K=1$. 
}
\label{fig4}
\end{figure}

%%%%%%%%%%%%%%%%%%%%%%%%%%%%
%% Figure 3
%%%%%%%%%%%%%%%%%%%%%%%%%%%%
\begin{figure}[t!]
\centering
\includegraphics[width=7cm]{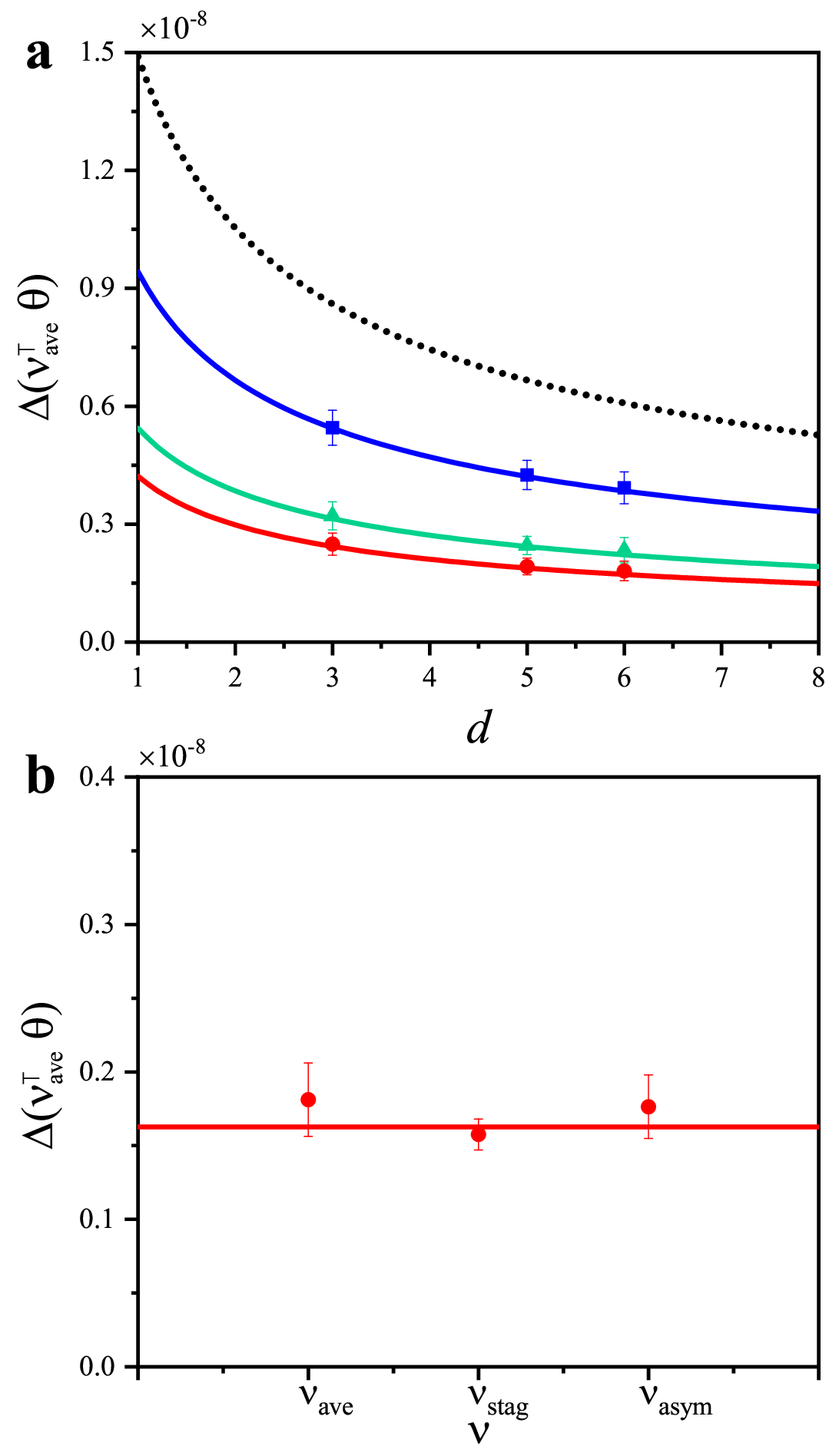}
\caption{\textbf{Demonstration of scalable distributed quantum sensing.}
\textbf{a}, Sensitivity $\Delta(\bm{\nu}_{\rm ave}^\top \bm{\theta})$ versus the number of sensors $d$. The average coherent-state photon number per MZI is fixed at $\bar{n}_c' = (4.5 \pm 0.13) \times 10^{15}$, and the squeezed-state photon number at $\bar{n}_s = 0.68 \pm 0.01$.
Sensitivity follows the predicted $1/\sqrt{d}$ scaling from Eq.~(\ref{sensEnt}) with $\Lambda = 0.14$. Symbols correspond to $K=1$ (blue squares), $K=3$ (green triangles), and $K=5$ (red circles).
\textbf{b}, Sensitivity $\Delta(\bm{\nu}^\top \bm{\theta})$ for different linear combinations of phase shifts.
Red circles show experimental results for $\bm{\nu}_{\rm ave} = \{1,1,1,1,1,1\}/d$ (average signal), $\bm{\nu}_{\rm stag} = \{1,-1,1,-1,1,-1\}/d$ (staggered signal), and $\bm{\nu}_{\rm asym} = \{1,1,1,-1,-1,-1\}/d$ (antisymmetric signal). The solid line represents the theoretical prediction from Eq.~(\ref{sensEnt}) for $\bar{n}_s = 0.68 \pm 0.01$, $\bar{n}_c = (2.7 \pm 0.077) \times 10^{16}$, and $\Lambda = 0.14$.
}
\label{fig5}
\end{figure}
%%%%%%%%%%%%%%%%%%%%%%%%%%%%
%%%%%%%%%%%%%%%%%%%%%%%%%%%%
%%%%%%%%%%%%%%%%%%%%%%%%%%%%

A second operating regime investigated in this work involves low optical intensity, where we optimize the allocation between $\bar{n}_c$ and $\bar{n}_s$ under a fixed total photon budget $\bar{n}_T = \bar{n}_c + \bar{n}_s$.
This regime is especially relevant for applications such as biological sensing~\cite{TaylorPR2016}, where excessive light intensity can damage the sample. 
Unlike prior DQS implementations in the low-photon regime—based on Fock~\cite{CiminiAP2023}, GHZ~\cite{LiuNATPHOT2021, ZhaoPRX2021}, or generalized NOON states~\cite{HongNATCOMM2021, KimNATCOMM2024}—our approach is based on continuous-variable Gaussian states~\cite{11}.
Figure~\ref{fig4} illustrates the sensitivity of the optimized entangled network as a function of $\bar{n}_T$.
In the experiment, the coherent input beams are attenuated to low-power levels before being injected into the MZIs, which are simultaneously fed with the entangled squeezed modes. 
We explore the crossover between the SQL and the HL, in very good agreement with the theoretical prediction in the presence of losses.
Reaching a phase uncertainty scaling $\Delta(\vect{\nu}^\top\vect{\theta})\sim \bar{n}_T^{-2}$ would require decreasing the particle losses and increasing the number of particles in the squeezed state (here limited to $\bar{n}_s \lesssim 1$).
In terms of absolute gains, the sensitivity improvements over the SQL are approximately a factor of about 3.7 with respect to the single-pass and 1.6 (about 4 dB for the variance) with respect to the five-pass configurations.

Scaling up the number of sensors is a critical requirement for practical DQS. 
In our implementation, scalability is achieved by varying the number of output ports of the QC by using additional beam splitters,  which distributes the entangled resource to an increasing number of interferometers.
The input state remains a single squeezed-vacuum state.
The scalability of our entangled network is demonstrated in Fig.~\ref{fig5}(a). 
There, we consider the uncertainty associated with estimating the average phase shift $\bm{\nu}_{\rm ave}^\top \bm{\theta}$, while fixing both the squeezed-state photon number (equally divided among the $d$ modes) and the average coherent-state photon number per interferometer.
As predicted by Eq.~(\ref{sensEnt}), we observe that the joint measurement sensitivity improves with increasing $d$, due to the larger number of MZIs contributing to the estimation. Specifically, we find
$\Delta^2(\bm{\nu}_{\rm ave}^\top \bm{\theta}) = \frac{e^{-2r} + \Lambda}{K d \bar{n}_c'}$,
where $\bar{n}_c' = \bar{n}_c / d$ is the average photon number per coherent mode.
Comparing results for $d = 3$, 5, and 6 interferometers confirms the expected scaling $\Delta(\bm{\nu}_{\rm ave}^\top \bm{\theta}) \sim 1/\sqrt{d}$ (colored solid lines), validating the scalability of our architecture relative to the SQL (black dotted line).

Finally, we examine the robustness of the network when estimating different combinations of phases, as defined in Eq.~(\ref{combination}). 
Up to this point, we focused on $\bm{\nu}_{\rm ave}$ corresponding to uniform averaging.
However, Eq.~(\ref{sensEnt}) predicts that equivalent sensitivity can be achieved for other signal structures, such as staggered phase shifts $\bm{\nu}_{\rm stag} = \{1,-1,1,-1,1,-1\}/d$ or antisymmetric signals $\bm{\nu}_{\rm asym} = \{1,1,1,-1,-1,-1\}/d$, upon optimizing the overall probe state.
In practice, implementing negative $\nu_j$ values requires a $\pi$ phase shift in the coherent input state $\ket{\alpha_j}$ with respect to the corresponding mode of the squeezed-vacuum state emerging from the quantum circuit.
This prediction is experimentally verified in Fig.~\ref{fig5}(b), where we present the sensitivity for three representative configurations of $\bm{\nu}$, all under the same $\bar{n}_c$, $\bar{n}_s$, and loss parameter $\Lambda$. 
Within experimental uncertainties, the results confirm that the optimal sensitivity is independent of the sign structure of $\bm{\nu}$.
This flexibility suggests that the optimized entangled network is suitable not only for spatially uniform signals, but also for estimating staggered or parity-modulated signal profiles.\\

\textbf{Discussion and Conclusions.} \\

We have demonstrated a scalable architecture for distributed phase sensing based on a network of MZIs.
By coherently distributing a single squeezed-vacuum state across multiple optical modes, and independently controlling the coherent amplitudes in each MZI, we realize collective phase estimation with quantum-enhanced precision. 
The use of multipass phase amplification further boosts sensitivity and resilience to photon losses. 
The key novelty of our approach is the exploitation of MZIs, which benefit from common-mode phase noise cancellation and thus guarantee noise resilience beyond homodyne phase sensing~\cite{GuoNATPHYS2020} and quadrature sensing~\cite{XiaPRL2020}.
A crucial advantage of our platform is its versatility -- addressing both low and high photon intensity regimes, which are relevant to different technological applications.
The architecture allows for real-time reconfiguration to measure arbitrary linear combinations of local phases. 
This flexibility is critical for practical sensing scenarios, such as field mapping, gradient estimation, or differential phase tracking across spatially distributed sensors. 
Moreover, the optimal allocation of quantum and classical resources enables our scheme to approach the QCRB using only local homodyne detection at a single output port of each MZI.

Our approach provides a new paradigm for quantum-enhanced metrology for a broad spectrum of applications.
In atomic clocks, for example, networks of entangled sensors can enable synchronized frequency measurements across distant nodes with precision beyond the SQL. 
Similarly, in cold-atom interferometry, distributed schemes based on shared squeezing can enhance the detection of gravitational gradients, rotations, or fundamental constants. 
Our architecture is naturally compatible with optical atomic systems~\cite{MaliaNATURE2022, CorgierQUANTUM2023, LiARXIV}, where local MZIs can be integrated with atomic ensembles or cavities acting as phase-sensitive elements. \\

\textbf{Methods.} \\

{\it Experimental setup.}
The experimental setup of a six-mode interferometer network is demonstrated in Fig. 2 (a) of main text. The mode
entanglement is generated with the minimum quantum resource of a single optical parametric amplifier (OPA),
and distributed to the Mach-Zehnder interferometers (MZIs). A Ti: sapphire laser (coherent MBR-110) with an
output power of about 2.2 W pumped by a green laser (Yuguang DPSS FG-VIIIB)
outputs coherent state of optical mode with wavelength at 895 nm. Besides
the injection mode of second harmonic generation (SHG) for the OPA pump, the
coherent laser is filtered by the mode cleaner for entangled MZI network,
including the signal mode of the OPA, the input coherent probe modes of the
MZIs and the local oscillator (LO) modes of balanced homodyne detections (BHDs). A bow-tie type OPA with
the type 0 quasi--phase-matched periodically poled KTiOPO4 (PPKTP) crystal
is used to generate the squeezed probe of 895nm under the pumping effect of
447 nm pump probe. The ring cavity consists of two planar mirrors and two
spherical mirrors (50 mm radius of curvature), as well as crystals with
dimensions of 1 mm $\times $ 2 mm $\times $ 10 mm. Two flat mirrors with
high reflection and a mode pair with a transmission of 5$\%$ at 895nm,
respectively, which are used as the input mirror and output mirror for the
OPA. And the other two flat concave mirrors are coated with a high
reflection film of 895 nm and an anti-reflection film of 447 nm. The seed
probe is incident from the flat mirror with high reflection. After the 447
nm pump probe generated by the SHG cavity is incident on the concave mirror
at 25 mW, it passes through the crystal in a single pass, and the nonlinear
crystal undergoes parametric down conversion into a squeezed state at 895
nm. With the relative phase of the pump probe and the seed probe to be 0, the
phase quadrature squeezed state is generated. This bow-tie shaped cavity is
actively scanned for cavity length through the 
, and then resonances are
locked using injected seed beams by Pound-Drever-Hall technology. The squeezing is distributed to six modes with of 5.30 $\pm $ 0.10 dB beyond standard quantum limit (SQL) by using QC. 

In each MZI, a coherent state probe and multiple pass interactions are
utilized to enhance the phase signals. Six coherent states are injected into
the MZIs and driven by the 4 MHz sinusoidal signals through the piezoelectric transducers (PZTs) to generate the phase changes. Furthermore,
the multiple pass interactions are implemented with two high-reflection
mirrors, the back-and-forth interaction greatly improves the phase signal. The phase
change is simulated by applying the 4 MHZ signal from 
PZT$_{j}$ at the
probe in the arm of each MZI, and multiple pass interactions are implemented
between two mirrors. 
The optical mode carrying phase signal in one arm of MZI and reference
optical mode in the other arm are interfered, and the destructive
interference output ports are detected, which avoids the saturation of
photo-diodes. The displacement sensitivity driven by PZTs is 8.9$\times $10$^{-16}$ m$\cdot$Hz$^{-1/2}$ with the total input coherent power
of 100 $\mu W$.

The BHDs are used to measure the quadrature phases of output modes of MZIs.
In each BHD measurement, the probe interferes with a LO optical mode to
obtain the quadrature components, and the resulting photocurrents by BHDs
are processed to derive the phase changes. The phase quadrature noise power
of destructive interference output port of the MZI is detected by the BHD by
locking the relative phase between the LO light and probe light, which overcomes
the problem of power saturation. The sensitivity with equal weight can be
obtained by jointly processing six BHD results. Our experimental system loss
and efficiency are: the quantum efficiency of the diode is 98$\%$, and the
other interference efficiency is 91$\%$. Therefore, the overall efficiency
of the entangled MZI network is 89$\%$.

For the stable locking of the cavity resonance and interference, the
reference optical beams are employed. A chopper in front of the OPA cavity
is used for switching the signal mode to the desired vacuum state, ensuring system stability. The output mode from the OPA is divided into six
entangled modes by a half-wave plate and a polarization
beam splitter (PBS) network. In addition, the
other input end of the interferometer uses an acousto-optic modulator (AOM)
to control the laser into a corresponding coherent state pulse. The relative phases between the squeezed and coherent
input modes in the MZI are locked by means of six high-gain detectors,
respectively. After the probe of the interferometer is combined by a 50:50
linear beam splitter (BS), the stable phase of the interferometer is locked by the other six
detectors. \\

{\it Discussion of theoretical phase sensitivities.}
We provide here further details on the theoretical methods and equations.
First, we have that $\bar{n}_c = \sum_{j=1}^d \vert \alpha_j \vert^2$ is the total mean photon number of the $d$ coherent states and $\bar{n}_s =\sinh^2 r$ is the average number of photons in the squeezed vacuum:
$\bar{n}_T = \bar{n
}_c + \bar{n}_s$ is the total average number of photons in the input probe state.
% 
%Notice that the efficiency in QC is near perfect, while $\eta_{dis}$ is changed to mimic fiber with different length.
%
For a given $\bm{\nu}$, the optimal photon allocation across the QC is achieved by selecting $P_j = \vert \alpha_j\vert^2/\bar{n}_c = \vert \nu_j \vert$, where $P_j$ is the probability that a photon of the squeezed vacuum state exits the $j$th port of the QC. 
These probabilities are implemented by
configuring the splitting performed by the QC~\cite{ReckPRL1994, ClementsOPTICA2016}.

The limits $\bar{n}_T\gg 1$ in Eq.~(\ref{optsqu}) is obtained from 
\begin{equation} \label{minrsensitivity}
\min_r~\Delta^2 (\bm{\nu}^\top \bm{\theta})_{\mathrm{opt}} = \frac{\big( 1 + 
\sqrt{1+4\Lambda\bar{n}_T}\big)^2}{4 K \bar{n}_T^2},
\end{equation}
which is the exact expression valid in all regimes (see \cite{supp} for details). 
In the ideal lossless case $\Lambda=0$, Eq.~(\ref{optsqu}) recovers the HL $1/(K\bar{n}_T^2)$, achieved when half of the $\bar{n}_T$ particles are in the squeezed-vacuum state and the other half are in the $d$ coherent states, i.e. $\bar{n}_s = \bar{n}_c=\bar{n}_T/2$. 
%
%In the presence of losses ($\Lambda > 0$), the sensitivity interpolates between the HL, for $\bar{n}_T \lesssim 1/\Lambda$, and a SQL-like scaling $\Lambda/(K \bar{n}_T )$, for $\bar{n}_T \gtrsim 1/\Lambda$. 
%
%Notably, a sub-SQL sensitivity is recovered asymptotically in $\bar{n}_T$ provided that $\eta>1/2$. 
%
%In the opposite regime, $\bar{n}_T \ll 1$, we find $\min_r~\Delta^2 (\bm{\nu}^\top \bm{\theta})_{\mathrm{opt}} = (1+\Lambda)/\bar{n}_T$ with the minimum achieved for $\bar{n}_s \ll \bar{n}_c$.

Finally, we recall that QCRB for our setup is~\cite{PezzeARXIV, supp} 
\begin{equation}
\Delta^2 (\bm{\nu}^\top \bm{\theta})_{\mathrm{QCRB}} = \frac{1}{K( \bar{n}_c
	e^{2r} + \sinh^2 r)}.
\end{equation}
In the regime $\bar{n}_c e^{2r} \gg \sinh^2 r$, it follows that $(\Delta \bm{\nu}^\top \bm{\theta})^2_{\mathrm{opt}} \approx (\Delta \bm{\nu}%
^\top \bm{\theta})^2_{\mathrm{QCRB}}$. \\

\textbf{Acknowledgements.} \\

This research was supported by the Innovation Program for Quantum Science and Technology (Grant No.2024ZD0300900), the Key Project of the National Key R \& D program of China (Grant No.2022YFA1404500), National Natural Science Foundation of China (Grant No.62122044, Grants, No.61925503, and No.62135008), the Fundamental Research Program of Shanxi Province (Grant No.202403021223001), and the fund for Shanxi “1331 Project” Key Subjects Construction.
LP acknowledge supports from the QuantERA project SQUEIS (Squeezing enhanced inertial sensing), funded by the European Union’s Horizon Europe Program and the Agence Nationale de la Recherche (ANR-22-QUA2-0006).

\newpage

\begin{widetext}
\begin{center}
    {\bf Supplemental Material} \\
\end{center}

In the following, we derive all equations discussed in the main text.
The derivation partially follows that of Ref.~\cite{PezzeARXIV2025}, which considered number counting measurements at each output port of the MZIs, rather than BHD considered in this work.
For completeness we repeat part of the derivation here.
For clarity, Fig.~\ref{Figure4SI}(a) and (b) show the entangled and separable, respectively, sensor network for the case $d=2$, clarifying the different schemes and the notation used in the main text.
We note that below, we do not impose the normalization of the vector $\vect{v}$. 
Equations of the main text are recovered when imposing $\sum_{j=1}^d \vert v_j \vert =1$.

%%%%%%%%%%%%%
% FIG 1SI
%%%%%%%%%%%%%
\begin{figure}[h!]
\includegraphics[width=0.75\columnwidth]{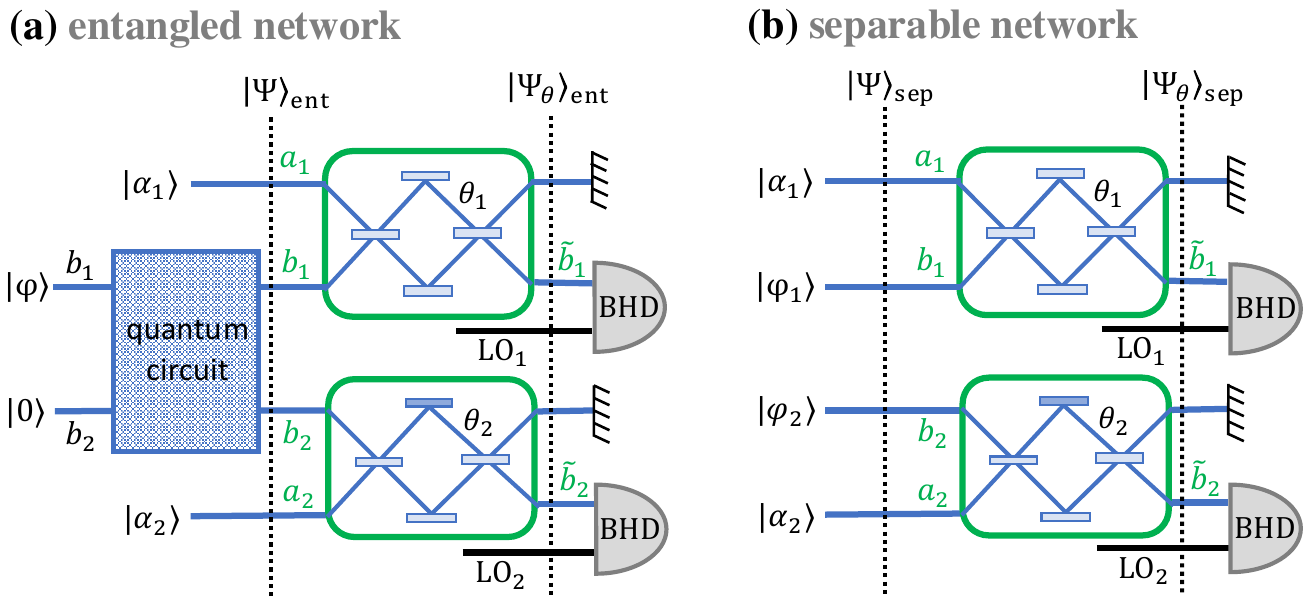}
\caption{Entangled (a) and separable (b) networks for the estimation of the relative phase shifts $\theta_1$ and $\theta_2$ in the two MZIs.
The vertical dotted line indicate the stage at which the quantum states $\ket{\Psi}$ and $\ket{\Psi_{\vect{\theta}}}$ are defined.
BHD indicates balance homodyne detection with LO (black line) being a local oscillator. 
} %%\tag{SuppFig.1}
\label{Figure4SI} 
\end{figure}
%%%%%%%%%%%%%
%%%%%%%%%%%%%
%%%%%%%%%%%%%

{\bf Probe state.}
We consider a $d$-mode quantum circuit (QC) with one input given by a generic single-mode state $\ket{\varphi} = \sum_{m=0}^{+\infty} c(m) \ket{m}$, where $c(m)$ are complex coefficients, and the other $d-1$ modes in the the vacuum.
Without loss of generality we consider the mode $b_1$ as the input mode of QC where the state $\ket{\varphi}$ is injected, see Fig.~\ref{Figure4SI}(a).
Below, we will also indicate this mode as $b$. 
Let us first show that the QC -- here represented by the unitary transformation $\hat{U}_{\rm QC}$ performs a multinomial splitting of the state $\ket{\varphi}$.
We have 
\beq
\ket{\Psi_{\rm QC}} &=& \hat{U}_{\rm QC} \big[ \ket{\varphi}\otimes \ket{0} \otimes ...  \otimes \ket{0} \big]  \nonumber \\ 
&=& \sum_{m=0}^{+\infty} c(m) \, \hat{U}_{\rm QC}^\dag \big[ \ket{m}\otimes \ket{0} \otimes ...  \otimes \ket{0} \big] \nonumber  \\
&=&
\sum_{m=0}^{+\infty} c(m) \, \frac{\hat{U}_{\rm QC}^\dag (\hat{b}_1^\dag)^m}{\sqrt{m!}} \big[ \ket{0}\otimes  \ket{0} \otimes ...  \otimes \ket{0} \big] \nonumber \\
&=& \sum_{m=0}^{+\infty} c(m) \, \frac{\hat{U}_{\rm QC}^\dag (\hat{b}_1^\dag)^m \hat{U}_{\rm QC}}{\sqrt{m!}} \big[ \ket{0}\otimes  \ket{0} \otimes ...  \otimes \ket{0} \big] \nonumber \\
&=& \sum_{m=0}^{+\infty} c(m) \, \frac{\big( \hat{U}_{\rm QC}^\dag \hat{b}_1^\dag\hat{U}_{\rm QC} \big)^m}{\sqrt{m!}}  \big[ \ket{0}\otimes  \ket{0} \otimes ...  \otimes \ket{0} \big], \nonumber 
\eeq
where we have used, respectively, $\ket{m} = \tfrac{(\hat{b}^\dag)^m \ket{0}}{\sqrt{m!}}$, $\hat{U}_{\rm QC} \big[ \ket{0}\otimes  \ket{0} \otimes ...  \otimes \ket{0} \big] = \ket{0}\otimes  \ket{0} \otimes ...  \otimes \ket{0}$, and the unitary properties of $\hat{U}_{\rm QC}\hat{U}_{\rm QC}^\dag = \Eins$.  
The QC performs a linear transformation of the $d$ modes: $\hat{U}_{\rm QC}^\dag \hat{b}_j^\dag \hat{U}_{\rm QC} = \sum_{k=1}^d (U_{\rm QC})_{j,k} \hat{b}_k^{\dag}$, where $U_{\rm QC}$ is a $d \times d$ unitary matrix. 
Restricting to mode transformations with real coefficients, we write 
\be %\tag{SM.1} 
\hat{U}_{\rm QC}^\dag \hat{b}_1^\dag \hat{U}_{\rm QC} = \sqrt{P_1} \hat{b}_1^\dag + \sqrt{P_2} \hat{b}_2^\dag + ... + \sqrt{P_d} \hat{b}_d^\dag,
\ee
where $P_j$ are real numbers with $\sum_{j=1}^d P_j = 1$ due to the conservation of atom number. 
The commutativity of different $\hat{b}_j$ allows to apply the multinomial theorem:
\be %\tag{SM.2}
\big( \hat{U}_{\rm QC}^\dag \hat{b}_1^\dag \hat{U}_{\rm QC} \big)^m = 
\big( \sqrt{P_1} \hat{b}_1^\dag + \sqrt{P_2} \hat{b}_2^\dag + ... + \sqrt{P_d} \hat{b}_d^\dag \big)^m = 
\sum_{\substack{m_1, ..., m_d =0 \\ m_1+ ... + m_d = m}}^{m} 
\frac{m!}{m_1! ... m_d!} \prod_{j=1}^m \big(\sqrt{P_j} \hat{b}_j^\dag \big)^{m_j}.
\ee
When applied to the vacuum, taking into account that $(\hat{b}_j)^{m_j}\ket{0} = \sqrt{m_j!} \ket{m_j}$, we have 
\be %\tag{SM.3}
\frac{\big( \hat{U}_{\rm QC}^\dag \hat{b}_1^\dag \hat{U}_{\rm QC} \big)^m}{\sqrt{m!}} \big[ \ket{0}\otimes  \ket{0} \otimes ...  \otimes \ket{0} \big] = \sum_{\substack{m_1, ..., m_d =0 \\ m_1+ ... + m_d = m}}^{m} 
\sqrt{\frac{m!}{m_1! ... m_d!}} P_1^{m_1/2} ... P_d^{m_d/2} \ket{m_1} ... \ket{m_d}. 
\ee
Finally, taking into account the sum over coefficients $c(m)$, we find that the output state of the QC is
\be \label{EQ.PsiQC} %\tag{SM.4}
\ket{\Psi_{\rm QC}} = \hat{U}_{\rm QC} \big[ \ket{\varphi}\otimes \ket{0} \otimes ...  \otimes \ket{0} \big] = \sum_{m=0}^{+\infty} c(m) \sum_{\substack{m_1, ..., m_d =0 \\ m_1+ ... + m_d = m}}^{m} 
\sqrt{\frac{m!}{m_1! ... m_d!}} P_1^{m_1/2} ... P_d^{m_d/2} \ket{m_1} ... \ket{m_d},
\ee 
giving
\be \label{cQC} %\tag{SM.5}
c_{\rm QC}(m_1, ..., m_d) = \sum_{m=0}^{+\infty} c(m)
\sqrt{\frac{m!}{m_1! ... m_d!}} P_1^{m_1/2} ... P_d^{m_d/2} \, \delta_{m_1+...+m_d,m},
\ee
where 
\be \label{Eq.Pj} %\tag{SM.6}
P_j = \frac{\bra{\Psi_{\rm QC}} \hat{b}_j^\dag \hat{b}_j \ket{\Psi_{\rm QC}}}{\sum_{j=1}^d \bra{\Psi_{\rm QC}} \hat{b}_j^\dag \hat{b}_j \ket{\Psi_{\rm QC}}},
\ee
is the probability to find a particle at the $j$th output mode of the QC.
We emphasize that the multinomial form of Eq.~(\ref{EQ.PsiQC}) holds since only one of the input modes of the QC is populated. 
If more than one input mode is not empty, then the form of $\ket{\Psi_{\rm QC}}$ is generally more involved.  

Let us indicate with $a_j$ and $b_j$ the two modes of the $j$th MZI, see Fig.~\ref{Figure4SI}(a).
The mode $a_j$ is in a coherent state $\ket{\alpha_j}$, with $\alpha_j = \vert \alpha_j \vert e^{i \phi_j}$, while the mode $b_j$ is one of the outputs of the QC.
The overall input state of the sensor network of $d$ MZIs is given by 
\be %\tag{SM.7}
\ket{\Psi}_{\rm ent} = \ket{\Psi_{\rm QC}} \otimes \ket{\vect{\alpha}},
\ee
where $\ket{\vect{\alpha}} = \ket{\alpha_1} \otimes ... \otimes \ket{\alpha_d}$.
The output state of the network is 
\be %\tag{SM.8}
\ket{\Psi_{\vect{\theta}}}_{\rm ent} = \bigotimes_{j=1}^d e^{-i \theta_j \hat{H}_j} \big( \ket{\Psi_{\rm QC}} \otimes \ket{\vect{\alpha}} \big),
\ee
where $\hat{H}_j  = (\hat{a}^\dag \hat{b} - \hat{b}^\dag \hat{a})/(2i)$ is the Hamiltonian of the $j$th interferometer and $\theta_j$ is the phase acquired in the $j$th MZI. \\

{\bf BHD measurements and phase sensitivity.}
We measure the quadrature operator $\tilde{q}_j = \tilde{b}_j + \tilde{b}_j^\dag$ in one output mode of the $j$th MZI, where $\tilde{b}_j$ and $\tilde{b}_j^\dag$ are annihilation and creation operators, respectively, for the output mode $b_j$, see Fig.~\ref{Figure4SI}(a).
We can write $\hat{q}_j$ as a function of quadrature operators in input, $\hat{q}_{b,j} = \hat{b}_j + \hat{b}_j^\dag$ and $\hat{q}_{a,j} =\hat{a}_j + \hat{a}_j^\dag$: 
\be \label{pideal} %\tag{SM.9}
\tilde{q}_{j} = 
e^{i \hat{H}_j \theta_j} \hat{q}_{b,j} e^{-i \hat{H}_j \theta_j} = \hat{q}_{b,j} \cos\frac{\theta_j}{2} + \hat{q}_{a,j} \sin\frac{\theta_j}{2}. 
\ee
Furthermore, we model the losses as a beam splitter with transitivity $\eta$.
The quadrature operator Eq.~(\ref{pideal}) thus generalizes to
\be \label{pout}  %\tag{SM.10}
\tilde{q}_{j}^{\eta} = \big( \sqrt{\eta}~ \hat{q}_{b,j} + \sqrt{1-\eta}~ \hat{q}_{0,j} \big) \cos\frac{\theta_j}{2} + \big( \sqrt{\eta}~\hat{q}_{a,j} + \sqrt{1-\eta}~\hat{q}_{0,j} \big) \sin\frac{\theta_j}{2}.  
\ee

Using the properties of the multinomial distribution, we calculate 
\begin{subequations} \label{bb}
\beq
&& \bra{\Psi_{\rm QC}} \hat{b}_j \ket{\Psi_{\rm QC}} = \sqrt{P_j}~\bra{\varphi} \hat{b} \ket{\varphi} \label{bb1} \\
&& \bra{\Psi_{\rm QC}} \hat{b}_j^\dag \hat{b}_j \ket{\Psi_{\rm QC}} = P_j~\bar{n}, \label{bb2} \\
&& \bra{\Psi_{\rm QC}} \hat{b}_j^\dag \hat{b}_k \ket{\Psi_{\rm QC}} = \bra{\Psi_{\rm QC}} \hat{b}_k^\dag \hat{b}_j \ket{\Psi_{\rm QC}} = \sqrt{P_j} \sqrt{P_k}~\bar{n}, \label{bb3} \\
&& \bra{\Psi_{\rm QC}} \hat{b}_j \hat{b}_k \ket{\Psi_{\rm QC}} = %\bra{\Psi_{\rm QC}} \hat{b}_k^\dag \hat{b}_j^\dag \ket{\Psi_{\rm QC}} = 
\sqrt{P_j} \sqrt{P_k}~\sum_{m=0}^{+\infty} c^*(m)c(m+2)\sqrt{(m+2)(m+1)} = \sqrt{P_j} \sqrt{P_k}~\bra{\varphi} \hat{b} \hat{b} \ket{\varphi}, \label{bb4}
\eeq
\end{subequations}
where $\bar{n} = \bra{\varphi} \hat{b}^\dag \hat{b} \ket{\varphi}$ is the average number of particles in the state $\ket{\varphi}$.
From Eq.~(\ref{bb1}), we thus find 
\be
\bra{\Psi_{\rm QC}} \hat{q}_{b,j} \ket{\Psi_{\rm QC}} = \sqrt{P_j}  \bra{\varphi}\hat{b}+\hat{b}^\dag \ket{\varphi}.
\ee
For $\ket{\varphi}$ such as the squeezed-vacuum or the Fock states, we find $\bra{\Psi_{\rm QC}} \hat{q}_{b,j} \ket{\Psi_{\rm QC}}=0$.
For the vacuum we also have $\bra{0} \hat{q}_{0,j} \ket{0} = 0$.
Therefore, the mean value of the output quadrature operator Eq.~(\ref{pout}) is 
\be
\mean{\tilde{q}_{j}^{\eta} } =
2 \sqrt{\eta}~\vert \alpha_j \vert \cos \phi_j  \sin\frac{\theta_j}{2},
\ee
and thus
\be \label{C}
\vect{C}_{j,k} = \frac{\partial \mean{\tilde{q}_{j}^{\eta} }}{\partial \theta_k} = \sqrt{\eta}~\vert \alpha_j \vert \cos \phi_j  \cos\frac{\theta_j}{2} ~\delta_{jk},
\ee
where $\delta_{jk}$ is the Dirac delta.
The correlations $\vect{\Gamma}_{j,k} = \mean{ \tilde{q}_{j}^{\eta} \tilde{q}_{k}^{\eta} } - \mean{ \tilde{q}_{j}^{\eta} } \mean{ \tilde{q}_{k}^{\eta} }$ between output quadrature operators are
\beq
\vect{\Gamma}_{j,k} =
\eta \bra{\Psi_{\rm QC}} \hat{q}_{b,j} \hat{q}_{b,k} \ket{\Psi_{\rm QC}} \cos\frac{\theta_j}{2} \cos\frac{\theta_k}{2} + 
\eta \bigg( 1 + 4 \vert \alpha_j \vert^2 \cos^2 \phi_j \bigg) \sin^2\frac{\theta_j}{2}~\delta_{j,k} + (1-\eta) \big( 1 + \sin \theta_j \big) ~\delta_{jk},
\eeq
where the second term is the contribution of the coherent states and the third terms comes from the losses, $\bra{0} \hat{q}_{0,j} \hat{q}_{0,k} \ket{0} = \delta_{jk}$.
In the following we restrict to the optimal working conditions $\phi_j = 0$ (corresponding to each coherent state being real) and $\theta_j=0$.
These conditions guarantee the maximum derivative Eq.~(\ref{C}).
We also consider the state $\ket{\varphi}$ to be real [namely, having real coefficients $c(m)$].
In particular, if $\ket{\varphi}$ is the squeezed-vacuum state, this assumption corresponds to this state having a null phase.  
To compute the first term, we use Eq.~(\ref{bb})
\beq
\bra{\Psi_{\rm QC}} \hat{q}_{b,j} \hat{q}_{b,k} \ket{\Psi_{\rm QC}} &=& \bra{\Psi_{\rm QC}} \hat{b}_j^\dag \hat{b}_k + \hat{b}_k^\dag \hat{b}_j  - \hat{b}_j \hat{b}_k -\hat{b}_j^\dag \hat{b}_k^\dag  \ket{\Psi_{\rm QC}} + \delta_{jk} \nonumber \\
&=& \sqrt{P_j} \sqrt{P_k} \bra{\varphi} 2 \hat{b}^\dag \hat{b} - \hat{b} \hat{b} - \hat{b}^\dag \hat{b}^\dag \ket{\varphi} + \delta_{jk} \nonumber \\
&=& \sqrt{P_j} \sqrt{P_k} \big[ (\Delta^2 \hat{p})_{\ket{\varphi}} - 1 \big] + \delta_{jk},
\eeq
For the working conditions discussed above, we finally arrive at 
\be
\vect{\Gamma}_{j,k} = \eta \sqrt{P_j} \sqrt{P_k} \big[ (\Delta^2 \hat{q})_{\ket{\varphi}} - 1 \big].
\ee
In this case we compute
\be
\vect{\Sigma}^{-1} = \vect{C}^{-1} \vect{\Gamma} \vect{C}^{-1} = \big[ (\Delta^2 \hat{p})_{\ket{\varphi}} - 1 \big] \vect{f} \vect{f}^\top + \vect{D}, 
\ee
where $\vect{f} = \{ \sqrt{P_1}/\vert \alpha_1\vert, ..., \sqrt{P_d}/\vert \alpha_d\vert \}$ and $\vect{D}$ is a diagonal matrix with elements $\vect{D}_{jj}=1/(\eta \vert \alpha_j\vert)$.
We thus arrive at the final expression for the phase uncertainty 
\be \label{Deltap}
\Delta^2 \vect{\nu}^\top \vect{\theta} = \vect{\nu}^\top \vect{\Sigma}^{-1} \vect{\nu} = \big[ (\Delta^2 \hat{q})_{\ket{\varphi}} - 1 \big] \Bigg( \sum_{j=1}^d \frac{\nu_j \sqrt{P_j}}{\vert \alpha_j \vert} \Bigg)^2 + \sum_{j=1}^d \frac{\nu_j^2}{\eta \vert \alpha_j \vert^2}.
\ee

{\bf Optimization of the sensor network.}
We want to optimize Eq.~(\ref{Deltap}) with respect to the intensities of the coherent state and the coefficients $P_j$ that characterize the QC.
First, we use the Cauchy-Schwartz inequality
\be
\Bigg( \sum_{j=1}^d \frac{\nu_j \sqrt{P_j}}{\vert \alpha_j \vert} \Bigg)^2 \leq \sum_{j=1}^d \frac{\nu_j^2}{\vert \alpha_j \vert^2},
\ee
giving 
\be \label{opt2}
\Delta^2 \vect{\nu}^\top \vect{\theta} \geq \frac{\eta \big[ (\Delta^2 \hat{q})_{\ket{\varphi}} - 1 \big] + 1 }{\eta} \Bigg( \sum_{j=1}^d \frac{\nu_j^2}{\vert \alpha_j \vert^2} \Bigg),
\ee
with equality if and only if 
\be \label{optPj}
P_j = \frac{\nu_j^2/\vert \alpha_j \vert^2 }{\sum_{j=1}^d \nu_j^2/\vert \alpha_j \vert^2}.
\ee
We further the right-hand side of Eq.~(\ref{opt2}) with respect to the coherent state intensities, for fixed $\bar{n}_c = \sum_{j=1}^d \vert \alpha_j \vert^2$. 
To this aim, we write the Lagrangian
\be 
\mathcal{L} = \frac{\eta \big[ (\Delta^2 \hat{q})_{\ket{\varphi}} - 1 \big] + 1 }{\eta} \Bigg( \sum_{j=1}^d \frac{\nu_j^2}{\vert \alpha_j \vert^2} \Bigg) + \lambda \sum_{j=1}^{d} \vert \alpha_j \vert^2, 
\ee
and compute $\partial \mathcal{L}/\partial \vert \alpha_j\vert^2=0$.
We find
\be \label{LM1} 
\lambda = \frac{\eta \big[ (\Delta^2 \hat{q})_{\ket{\varphi}} - 1 \big] + 1 }{\eta} \frac{\nu_j^2}{\vert \alpha_j\vert^4}.
\ee
In particular, we thus have that that $\vert \nu_j\vert$ is proportional to $\vert \alpha_j\vert^2$, namely
\be \label{optalpha}
\frac{\vert \alpha_j\vert^2}{\bar{n}_c} = \frac{\vert \nu_j \vert}{\sum_{j=1}^d \vert \nu_j \vert}.
\ee
Replacing the optimal conditions for $\vert \alpha_j \vert^2$, Eq.~(\ref{optalpha}), and for $P_j$, Eq.~(\ref{optPj}), into Eq.~(\ref{Deltap}), we find
\be \label{optsens1}
\Delta^2 (\vect{\nu}^\top \vect{\theta})_{\rm opt} = \frac{(\Delta^2 \hat{q})_{\ket{\varphi}} + \Lambda}{\bar{n}_c} \Bigg( \sum_{j=1}^d \vert \nu_j \vert \Bigg)^2.
\ee
where $\Lambda = 1/\eta - 1$.
After multipass interactions between the probe light and the sample, the phase uncertainty is enhanced to 
\be \label{optsens0}
\Delta^2 (\vect{\nu}^\top \vect{\theta})_{\rm opt} = \frac{(\Delta^2 \hat{q})_{\ket{\varphi}} + \Lambda}{K\bar{n}_c} \Bigg( \sum_{j=1}^d \vert \nu_j \vert \Bigg)^2,
\ee
where $K$ is the number of multipass interaction.
In the case of a squeezed-vacuum state, we have $(\Delta^2 \hat{q})_{\ket{\varphi}} = e^{-2r}$ and recover Eq.~(\ref{sensEnt}) (we recall that the normalization $\sum_{j=1}^d \vert \nu_j \vert=1$ is assumed in the main text). 
Notice that $0\leq \eta \leq 1$ such that $\Lambda \geq 0$: the noiseless case $\eta=1$ corresponds to $\Lambda=0$.
When $\Lambda > 0$, the contribution from $\Lambda$ in the numerato of Eq.~(\ref{optsens0}) may dominate over the squeezing term $e^{-2r}$: increasing the squeezing coefficient becomes ineffective to increase the sensitivity of the device. \\

{\bf Optimization of the squeezed state.}
We can further optimize Eq.~(\ref{optsens1}) with respect to the squeezing strength $r$, when fixing the loss coefficient $\eta$ and the total average number of particles $\bar{n}_T = \bar{n}_c + \bar{n}_s$, where $\bar{n}_s = \sinh^2 r$ and $(\Delta^2 \hat{q})_{\ket{\varphi}} = e^{-2r}$.
The relation between photon number of squeezed state and squeezing factor is  
\be \label{exp2r}
e^{-2r}=1+2\bar{n}_s-2\sqrt{\bar{n}_s+\bar{n}^2_s}.
\ee
Using Eq.~(\ref{exp2r}), we find that Eq.~(\ref{optsens0}) can be expressed as
\be \label{optsens2}
\Delta^2 (\vect{\nu}^\top \vect{\theta})_{\rm opt} = \frac{1+2\bar{n}_s-2\sqrt{\bar{n}_s+\bar{n}^2_s} + \Lambda}{K(\bar{n}_T - \bar{n}_s)} \Bigg( \sum_{j=1}^d \vert \nu_j \vert \Bigg)^2.
\ee
Results of the numerical optimization of Eq.~(\ref{optsens2}) are shown in Fig.~(\ref{Figure2SI}).
For $r \gg 1$ we can use $e^{-2r}\approx 1/(4\bar{n}_s)$ and perform an analytical optimization: we take the derivative of Eq.~(\ref{optsens2}) with respect to $\bar{n}_s$ and imposing $\partial \Delta^2 (\vect{\nu}^\top \vect{\theta})_{\rm opt}/\partial \bar{n}_s=0$.
We find that the optimal photon number in the squeezed state is
\be \label{squopt2}
\bar{n}_{s,{\rm opt}} 
%= \frac{1+4\bar{n}_T\eta-2\bar{n}_T\eta^2-\sqrt{(2\bar{n}_T\eta^2-4\bar{n}_T\eta-1)^2-4\bar{n}^2_T\eta^2(4\bar{n}_T\eta^2+\eta^2-4\bar{n}_T\eta-1)}}{2(\eta-1)(4\bar{n}_T\eta+\eta+1)} \approx =
\approx \frac{\bar{n}_T}{1+\sqrt{1+4\Lambda\bar{n}_T}}, \qquad {\rm for}\,\, \bar{n}_s\gg 1,
\ee
to the leading order.
In the noiseless case $\Lambda=0$, we find $\bar{n}_{s,{\rm opt}} = \bar{n}_T/2$: the optimal squeezing strength is obtained when half or the total average number of particles are in the squeezed state.
In the presence of losses, it is more convenient to have a smaller squeezing and to increase the number or particles in the coherent states. 
By replacing Eq.~(\ref{squopt2}) into Eq.~(\ref{optsens2}), we find 
\be \label{optsens5}
\min_{r} \Delta^2 (\vect{\nu}^\top \vect{\theta})_{\rm opt}  
%\frac{1+2\bar{n}_{s,{\rm opt}}-2\sqrt{\bar{n}_{s,{\rm opt}}+\bar{n}^2_{s,{\rm opt}}}+\Lambda}{K \bar{n}_{s,{\rm opt}}}\Bigg( \sum_{j=1}^d \vert \nu_j \vert \Bigg)^2 
\approx \frac{(1+\sqrt{1+4 \Lambda \bar{n}_T})^2}{4 K \bar{n}_T^2}\Bigg( \sum_{j=1}^d \vert \nu_j \vert \Bigg)^2, \qquad {\rm for}\,\, \bar{n}_s\gg 1.
\ee
In the opposite limit, we expand Eq.~(\ref{optsens2}) in Taylor series, 
\be
\Delta^2 (\vect{\nu}^\top \vect{\theta})_{\rm opt} = \Bigg( \frac{1+\Lambda}{\bar{n}_T} - \frac{2\sqrt{\bar{n}_s}}{\bar{n}_T} + O(\bar{n}_s) \Bigg) \Bigg( \sum_{j=1}^d \vert \nu_j \vert \Bigg)^2.
\ee
Taking the derivative with respect to $\bar{n}_s$ gives
\be
\bar{n}_{s,{\rm opt}} 
\approx \frac{\bar{n}_T^2}{(1+\Lambda)^2} + O(\bar{n}_T^3), \qquad {\rm for}\,\, \bar{n}_s\ll 1.
\ee
The optimal value of $\bar{n}_s$ has a leading second order contribution to $\bar{n}_T$.
We thus conclude that 
\be \label{optsens6}
\min_{r} \Delta^2 (\vect{\nu}^\top \vect{\theta})_{\rm opt}  
%\frac{1+2\bar{n}_{s,{\rm opt}}-2\sqrt{\bar{n}_{s,{\rm opt}}+\bar{n}^2_{s,{\rm opt}}}+\Lambda}{K \bar{n}_{s,{\rm opt}}}\Bigg( \sum_{j=1}^d \vert \nu_j \vert \Bigg)^2 
\approx \frac{1+\Lambda}{\bar{n}_T} \Bigg( \sum_{j=1}^d \vert \nu_j \vert \Bigg)^2 \qquad {\rm for}\,\, \bar{n}_s\ll 1.
\ee

%%%%%%%%%%%%%
% FIG 1SI
%%%%%%%%%%%%%
\begin{figure}[b!]
\includegraphics[width=1\columnwidth]{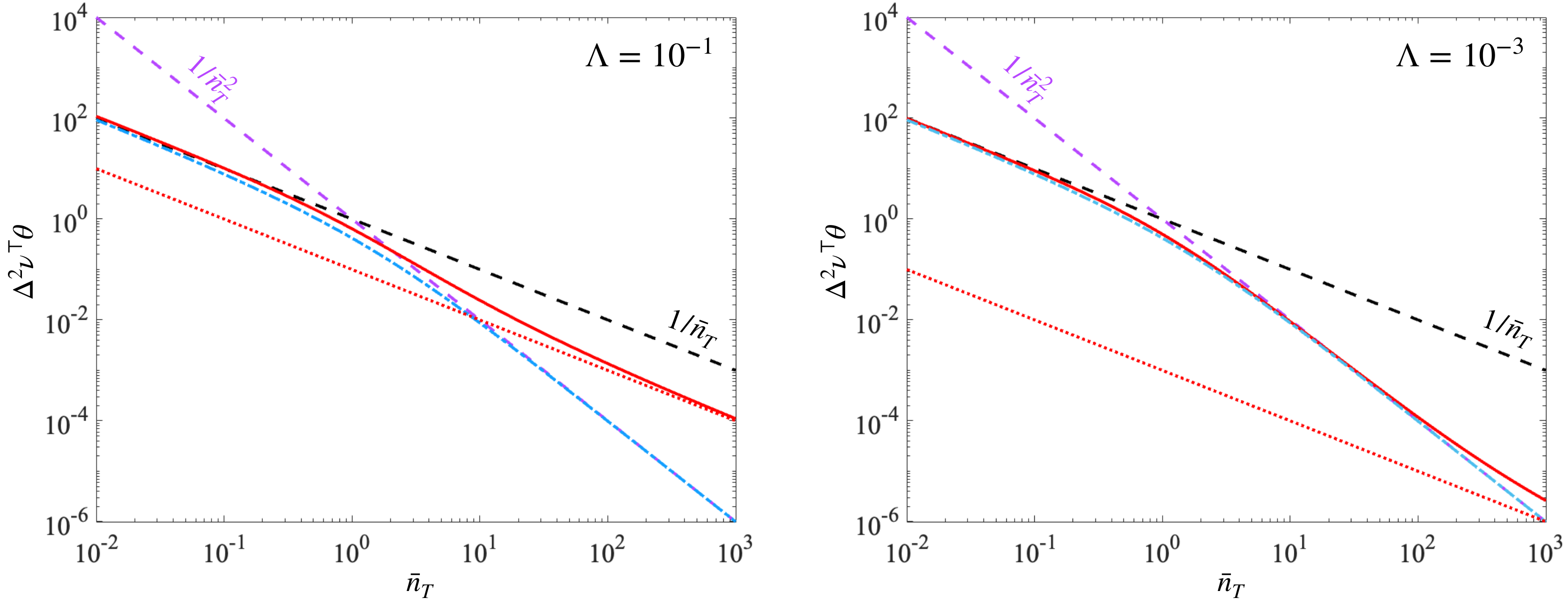}
\caption{Optimized phase variance as a function of $\bar{n}_T$, for $K=1$ and $\sum_{j=1}^d \vert \nu_j \vert=1$. 
The two panels refer to $\Lambda=10^{-1}$ (left) and $\Lambda=10^{-3}$ right.
In both panels the black dashed line $1/\bar{n}_T$; the magenta dashed line is $1/\bar{n}_T^2$; the solid red line is the numerical minimization of Eq.~(\ref{optsens2}); the dotter red line is the asymptotic limit $\Lambda/\bar{n}_T$; and the dot-dashed blue line is the numerical minimization of the QCRB.
} \label{Figure2SI} 
\end{figure}
%%%%%%%%%%%%%
%%%%%%%%%%%%%
%%%%%%%%%%%%%

{\bf Quantum Cram\'er-Rao bound.}
The quantum Cram\'er-Rao bound for the scheme of Fig.~\ref{Figure4SI}(b), in the lossless case, has been provided in Ref.~\cite{PezzeARXIV}. 
When optimized with respect to the coherent state intensities, it is given by 
\be \label{QCRB}
\Delta^2(\vect{v}\cdot\vect{\theta})_{\rm QCRB} = \frac{1}{\bar{n}_c (\Delta^2 \hat{p})_{\ket{\varphi}} + \bar{n}} \Bigg( \sum_{j=1}^d \vert \nu_j \vert \Bigg)^2,
\ee
where $\hat{p} = (\hat{b}-\hat{b}^\dag)/i$ (here $p$ is assumed to be the quadrature with the largest variance, see~\cite{PezzeARXIV}) and $\bar{n}$ is the average number of particles in the state $\ket{\varphi}$.
Equation~(\ref{QCRB}) should be compared with Eq.~(\ref{optsens1}) in the lossless case ($\eta=1$),
\be \label{optsenslossless}
\Delta^2 (\vect{\nu}^\top \vect{\theta})_{\rm opt} = \frac{(\Delta^2 \hat{q})_{\ket{\varphi}}}{\bar{n}_c} \Bigg( \sum_{j=1}^d \vert \nu_j \vert \Bigg)^2.
\ee
As a consequence of the Heisenberg uncertainty relation $(\Delta^2 \hat{q})_{\ket{\varphi}} (\Delta^2 \hat{p})_{\ket{\varphi}} \geq 1$, we have $\Delta^2(\vect{v}\cdot\vect{\theta})_{\rm QCRB} \leq (\Delta^2 \vect{\nu}^\top \vect{\theta})_{\rm opt}$.
For the squeezed-vacuum state, we have $\Delta^2( \hat{q})_{\ket{\varphi}} = e^{-2r}$, $\Delta^2 (\hat{q})_{\ket{\varphi}} = e^{2r}$ and $\bar{n} = \sinh^2r$.
We see that 
\be \label{equality}
\Delta^2 (\vect{\nu}^\top \vect{\theta})_{\rm opt} \approx \Delta^2(\vect{v}\cdot\vect{\theta})_{\rm QCRB}
\ee
for $\bar{n}_c e^{2r} \gg \sinh^2r$, which is verified in our experiment.
We notice that Eq.~(\ref{equality}) is verified for $\theta_j =0$. 
While $\Delta^2 (\vect{\nu}^\top \vect{\theta})_{\rm opt}$ increases with $\theta_j$, $\Delta^2(\vect{v}\cdot\vect{\theta})_{\rm QCRB}$ remains constant.

Using $e^{2r}=1+2\bar{n}_s +2\sqrt{\bar{n}_s+\bar{n}^2_s}$, we have that Eq.~(\ref{QCRB}) can be expressed as
\be 
\Delta^2 (\vect{\nu}^\top \vect{\theta})_{\rm opt} = \frac{1}{(\bar{n}_T - \bar{n}_s)(1+2\bar{n}_s+2\sqrt{\bar{n}_s+\bar{n}^2_s}) + \bar{n}_s}\Bigg( \sum_{j=1}^d \vert \nu_j \vert \Bigg)^2.
\ee
A numerical optimization of this equation with respect to $\bar{n}_s$, for fixed $\bar{n}_T$ is shown by the dot-dashed blue line in Fig.~\ref{Figure2SI}.
In the limit $\bar{n}_T\gg 1$, Eq.~(\ref{QCRB}) can be optimized with respect to $\bar{n}_s$, giving 
\be
\min_{r} \Delta^2 (\vect{\nu}^\top \vect{\theta})_{\rm QCRB} = \frac{1}{\bar{n}_T^2} \Bigg( \sum_{j=1}^d \vert \nu_j \vert \Bigg)^2, \qquad {\rm for}\,\, \bar{n}_s\gg 1.
\ee
with $\bar{n}_{s,{\rm opt}} = \bar{n}_{c,{\rm opt}} =\bar{n}_T/2$,
and 
\be2
\min_{r} \Delta^2 (\vect{\nu}^\top \vect{\theta})_{\rm QCRB} = \frac{1}{\bar{n}_T} \Bigg( \sum_{j=1}^d \vert \nu_j \vert \Bigg)^2, \qquad {\rm for}\,\, \bar{n}_s\ll 1.
\ee
These limits agree with Eqs.~(\ref{optsens5}) and~(\ref{optsens6}) when the latter are computed for $\Lambda=0$. \\

{\bf Separable sensor network.}
The separable sensor network is made of $d$ independent MZIs, see Fig.~\ref{Figure4SI}(b), with $\ket{\varphi_j}$ being the input of mode $b_j$, the other input being a coherent state $\ket{\alpha_j}$.
The overall probe state of the separable sensor network is 
\be
\ket{\Psi}_{\rm sep} =  \bigotimes_{j=1}^d \big( \ket{\varphi_j} \otimes \ket{\alpha_j} \big),
\ee
while the output state is 
\be
\ket{\Psi_{\vect{\theta}}}_{\rm sep} = \bigotimes_{j=1}^d e^{-i \hat{H}_j \theta_j}\big( \ket{\varphi_j} \otimes \ket{\alpha_j} \big).
\ee
Also in this case, we consider quadrature measurement at the output port of each MZI and compute the corresponding variance via error propagation
\be
\Delta^2 \theta_j = \frac{(\Delta \tilde{q}_j)^2}{(\partial \mean{\tilde{q}_j}/\partial \theta_j)^2} = \frac{(\Delta^2 \hat{q})_{\ket{\varphi_j}} + \Lambda}{\vert\alpha_j\vert^2}.
\ee
The sensitivity for the estimation of a linear combination of phases is 
\be \label{senssep}
\Delta^2 (\vect{\nu}^\top \vect{\theta})_{\rm sep} = \sum_{j=1}^d \nu_j^2 \Delta^2 \theta_j = \sum_{j=1}^d \nu_j^2 \frac{(\Delta^2 \hat{p})_{\ket{\varphi_j}} + \Lambda}{\vert\alpha_j\vert^2}.
\ee
We want to optimize Eq.~(\ref{senssep}) with respect to $\vert \alpha_j \vert^2$ for fixed $\bar{n}_c = \sum_{j=1}^d \vert \alpha_j \vert^2$.
Let us first assume that $\ket{\varphi_j} = \ket{\varphi}$.
In this case we have 
\be
\Delta^2 (\vect{\nu}^\top \vect{\theta})_{\rm sep} = \Big[ (\Delta^2 \hat{p})_{\ket{\varphi}} + \Lambda ] \sum_{j=1}^d \frac{\nu_j^2}{\vert \alpha_j\vert^2}.
\ee
The optimization is analogous to the one preformed previously and the optimal condition in provided by Eq.~(\ref{optalpha}), giving
\be 
\min_{\ket{\alpha_1}, ..., \ket{\alpha_d}} \Delta^2 (\vect{\nu}^\top \vect{\theta})_{\rm sep} = \frac{(\Delta^2 \hat{p})_{\ket{\varphi}} + \Lambda}{\bar{n}_c} \Bigg( \sum_{j=1}^d \vert \nu_j \vert \Bigg)^2,
\ee
which precisely coincides with Eq.~(\ref{optsens1}).
When fixing the state $\ket{\varphi}$ and the total intensity of the $d$ coherent states, the entangled network reaches the same sensitivity as the separable one, despite the fact that the latter uses $d$ copies of the state $\ket{\varphi}$, while the former consider only one copy of $\ket{\varphi}$ optimally split by the QC.

Let us now consider the case where each state $\ket{\varphi_j}$ can be different and let us optimize the full sensor network with respect to each $\vert \alpha_j \vert^2$ and $\ket{\varphi_j}$. 
This optimization is more complex and we only consider two interesting limits: we have 
\be
\min_{\ket{\alpha_1}, ..., \ket{\alpha_d}} \min_{\ket{\varphi_1}, ..., \ket{\varphi_j}} \Delta^2 (\vect{\nu}^\top \vect{\theta})_{\rm opt} = 
\begin{cases}
    \frac{1}{\bar{n}_T^2} \big( \sum_{j=1}^d \vert \nu_j \vert^{2/3} \big)^3, & {\rm for}~\bar{n}_s \lesssim \Lambda, \\
    \hspace{0.05cm} \\
    \frac{\Lambda}{\bar{n}_T}, & {\rm for}~\bar{n}_s \gtrsim \Lambda.
\end{cases}
\ee
We thus recover Eq.~(\ref{gain}) of the main text.

\end{widetext}

\end{document}